\documentclass[11pt,a4paper,twocolumn,notitlepage]{article}
\usepackage{amssymb,amsmath}
\usepackage{amsthm}
\usepackage{bm}
\usepackage[french]{babel}
\usepackage[latin1]{inputenc}
\usepackage[geometry,weather,misc,clock]{ifsym}
\usepackage{hyperref}
\hypersetup{pdftex, colorlinks=true, linkcolor=blue, citecolor=blue, filecolor=blue, urlcolor=blue, pdftitle=DWs, pdfauthor=vdolocan , pdfsubject=, pdfkeywords=}
\usepackage[pdftex]{graphicx}

\makeatletter

\begin{document}

\title{Metastable domain wall dynamics in magnetic nanowires}
\date{}
\maketitle

\author{Voicu O. Dolocan\footnote{ \emph{Email:} voicu.dolocan@im2np.fr} \\
Aix-Marseille University, Marseille, France \and IM2NP CNRS, Avenue Escadrille Normandie Niemen, 13397 Marseille, France
}

\abstract{
We study the formation and control of metastable states of pairs of domain walls in cylindrical nanowires of small diameter where the transverse walls are the lower energy state. We show that these pairs form bound states under certain conditions, with a lifetime as long as 200ns, and are stabilized by the influence of a spin polarized current. Their stability is analyzed with a model based on the magnetostatic interaction and by 3D micromagnetic simulations. The apparition of bound states could hinder the operation of devices.}

%


\section{Introduction}

Domain walls (DW) form a frontier region between adjacent ferromagnetic domains with different direction of magnetization. Their internal structure depends on the competition between the local exchange and anisotropy and the non-local dipolar field\cite{Hubert}. When reducing the size of the objects to the nanoscale, the shape starts to play an important role. Magnetic nano-materials as nanowires, nanotubes and nanodots can be designed to have specific properties and arrays of patterned nano-materials can be used as storage or logic devices. This devices are mainly based on DWs motion that can be controlled by magnetic field or electric current\cite{Tatara,Boulle}.

Recently, possibles applications were proposed based on the current induced motion of the domain walls in nanowires\cite{Allwood,Parkin,Hayashi}. Usually, transverse (TDW) or vortex walls (N\'{e}el type walls) appear in nanowires between head-to-head (HH) or tail-to-tail (TT) domains. Sometimes, these two wall types can transform one into another during propagation and even other types can emerge\cite{Klaui}. In small diameter cylindrical nanowires, the Walker limit is completely suppressed for TDWs with theirs velocity and precession speed depending linearly on the applied current\cite{Yan2}.

The interactions between DWs\cite{Hayward1,Hayward2} or between a DW and artificially patterned traps\cite{Petit,OBrien} are used to manipulate the propagation of DWs. Often, these interactions are pictured in a first approximation in the Columbian approach, introducing the magnetostatic charge (pole) density  $\rho = -\mu_0\nabla$\textbf{m} and surface charge density $\sigma = \textbf{m}\cdot\textbf{n}$. A HHDW carries an intrinsic North monopole moment or a positive charge while a TTDW carries a South monopole moment or negative charge\cite{Hayward1}. Like charges repel, while opposite charges attract. 

The interaction between DWs can also be interpreted using topology. A DW as a nonuniform magnetization configuration, can be viewed as a topological defect or soliton\cite{Braun}. A topological defect is characterized by the winding number \textit{n}, which counts the number of times the field wraps around the unit circle when crossing the defect.  The DWs in flat nanowires are composite objects formed of edge defects with half-integer winding numbers\cite{Tchernyshyov,Kunz} (\textit{n} = $\pm$1/2). Two edge defects of opposite winding number\cite{Tretiakov} belong to the same topological sector and can be deformed continuously into the ground state and annihilate (attract). Two defects of same winding number cannot be deformed continuously into a zero charge state and thus repel each other. An external force is needed to annihilate them.

In this letter, we study the interaction between DWs in cylindrical ferromagnetic nanowires of moderate aspect ratio. Understanding theirs behavior and the possible metastable states that can appear, as local equilibrium states, are important for the progress of devices. The stability of these states is discussed both in the free and in the forced regime. The free regime consists of the relaxation of DWs in the nanowire, while the forced regime comprises the interaction of DWs with a spin polarized electric current. We show that metastable DWs pairs are created in small diameter cylinders and can form oscillatory bound states depending on the local potential landscape. Under the influence of a DC current, the bound states can be moved along the nanowire in both directions. These pairs could assist in the reversal mechanism of small nanowires\cite{Braun2}.

\begin{figure}[!t]
\center
  \includegraphics[width=8cm]{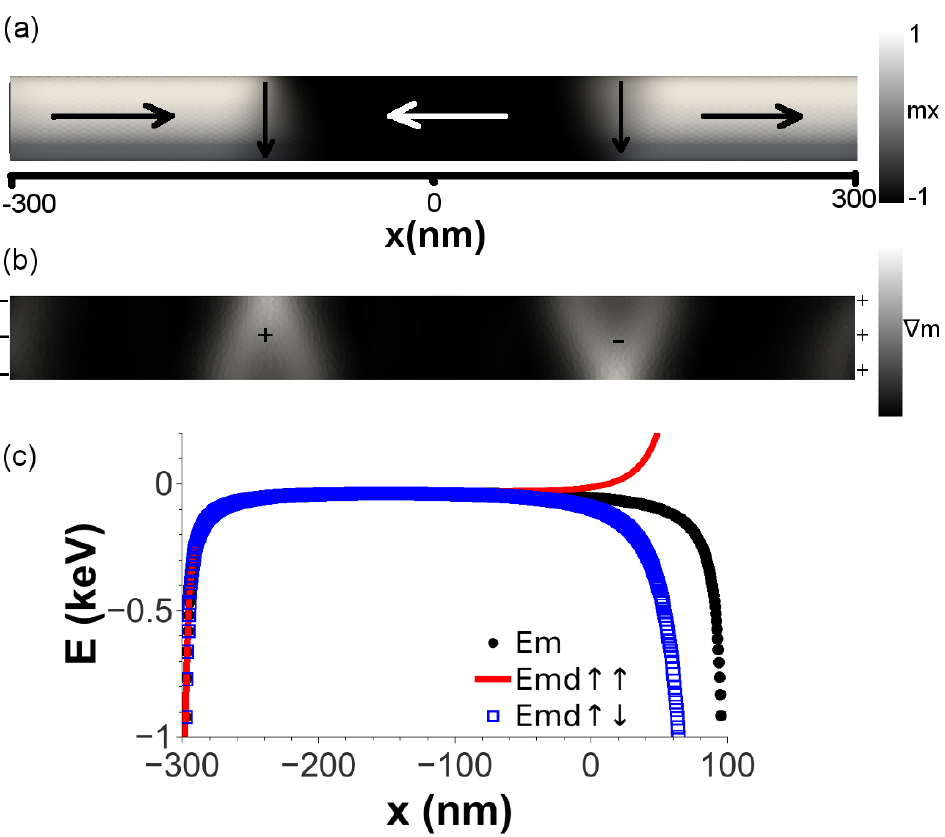}
 \caption{\label{Fig.1} (a) Two transverse domain walls with opposite chirality in a cylindrical nanowire of 600nm in length (longitudinal $x$ axis) and 60nm in diameter. (b) Magnetic density charge magnitude of two TDWs as in panel (a). The surface charge density at the edges is indicated along with the polarity of the DWs magnetic charges (poles). (c) Calculated potential landscape of a rigid DW, as in the panels (a)-(b), situated at the position $x$=-100nm between the edge of the nanowire (at $x$=-300nm) and another DW with opposite chirality (at $x$=+100nm). The filled circles correspond to the monopole-monopole interaction, while the full line and the empty squares correspond to the interaction energy with dipolar correction when the two DWs point in the same direction or opposite.}
\end{figure}

\section{One dimensional model}

The motion of a DW is usually described using the rigid 1D approximation\cite{Slonc,Thiaville}. This model gives a qualitative understanding of the motion of TDWs. The equations of motion of the DW are :

\begin{align}
\label{eq1}
(1+\alpha^2)\dot{X} =& -\frac{\alpha\gamma\Delta}{2\mu_0M_s S}\frac{\partial E}{\partial X} + \frac{\gamma\Delta}{2}H_k\sin 2\psi \nonumber\\
& + \frac{\gamma}{2\mu_0M_s S}\frac{\partial E}{\partial\psi} + (1+\alpha\beta)u \\ 
(1+\alpha^2)\dot{\psi} =& -\frac{\gamma}{2\mu_0M_s S}\frac{\partial E}{\partial X} -\frac{\gamma\alpha}{2}H_k\sin 2\psi \nonumber\\
& - \frac{\alpha\gamma}{2\Delta\mu_0 M_s S}\frac{\partial E}{\partial\psi} + \frac{\beta-\alpha}{\Delta}u
\end{align}

\noindent where $X$ and $\psi$ are the position and azimuthal angle (in the $yz$ plane) of the DW, $\Delta$ the DW width, $\gamma$ the gyromagnetic ratio, $M_s$ the saturation magnetization, $H_k$ the DW demagnetizing field, $\alpha$ the damping parameter, $u$ the spin drift velocity and $\beta$ the non-adiabatic parameter. $E$ is the potential energy of the DW that includes the internal energy, the Zeeman energy, the interaction energy with other DW and eventually the pinning energy.

The interaction energy between two DWs of opposite chirality in the same wire was calculated analytically using the point-charge model with multipole expansion\cite{Kruger}. The expression below disregards the internal structure of the DWs and the exchange interaction between them, considering only the interaction through the dipolar field. As long as the distance between the two DWs remains large (above three times the DW width), this approximates well the interaction of the DWs and gives a qualitative description of the interaction. In the numerical simulation of the next section, the exchange energy is taken into account and as will be shown in Fig.~\ref{Fig.2}(d) it's smaller than the demagnetizing energy by a factor 4 when the DWs are at large distances. 

In this 1D approximation the interaction energy between two DWs separated by $d$ is:

\begin{align}
\label{eq2}
E &= \frac{\mu_0q_1q_2}{4\pi d} \left( 1 - \frac{\pi^2\Delta^2\cos(\psi_1-\psi_2))}{4d^2} \right) \nonumber\\ &= -\frac{\mu_0M_s^2S^2}{\pi d}\left( 1 - \frac{\pi^2\Delta^2\cos(\psi_1-\psi_2))}{4d^2} \right)
\end{align}

Here $q_{ij}$ represent the 'magnetic charges' of the DWs and in this approximation are equal to $\pm2M_s$S, with S the section of the nanowire. The first term represents the Coulomb-type interaction energy (monopole-monopole), while the second is the correction from the dipole-dipole interaction. The next correction term (quadrupolar contribution) does not change the potential landscape calculated below. The interaction between the DWs depends mainly on the separation distance $d$.  The interaction energy increases with $d$, therefore one DW experiences an energy well created by the second DW\cite{Hayward1}. The correction term becomes important at moderate distances ($d \sim 3-5\Delta$) and depends on the orientation of the DWs dipolar moment. It takes extremal values when the DWs dipolar moment points in the same direction or opposite. This equation remains valid as long as the DWs do not overlap.

In a finite size nanowire, the charged DW also interacts with the edge charges (Fig.~\ref{Fig.1}(b)) that create another potential well with different depth. The monopoles of each end of the wire are given by $\pm M_sS$. The total interaction energy of a rigid DW is shown in Fig.~\ref{Fig.1}(c). The DW is situated at half distance between the edge of the nanowire (at the position $x$=-300nm) and another DW of opposite chirality (at the position $x$=+100nm), having a width $\Delta$ = 30nm. When considering only the first term in Eq.~\eqref{eq2} (filled cercles), the DW fells an attraction from the second DW as they have opposite charges. If the dipolar correction is taken into account, the DW is attracted or repelled by the other DW when theirs dipolar moments point in opposite or same direction (empty squares or full line) and this interaction decreases more slowly as in the monopolar case. As a result, if the magnetization in the two DWs precesses with different angular velocity, the interaction energy between them oscillates between attraction and repulsion leading to a metastable bound state. As the precession in TDWs is related with changes in position, this type of metastable bound state should oscillate along the symmetry axis of the nanowire.


\section{Numerical simulation}

To investigate the dynamics of DW pairs and the predictions of the analytical 1D model, we compute full 3D micromagnetic simulations (with the nmag package\cite{Fischbacher}) to determine the spatial distribution of the magnetization dynamics\cite{Dolocan1}. We use cylinders with a length varying between 600nm and 1200nm and diameters ranging from 10nm to 60nm. The cylinders were discretized into a mesh with a cell size between 0.9nm, for thinner cylinders, to 3nm for thickest, inferior to the exchange length ($\sim$5nm for Ni). We consistently checked that smaller cell size discretization does not influence the results presented below.

\begin{figure}[!t]
  \includegraphics[width=6cm]{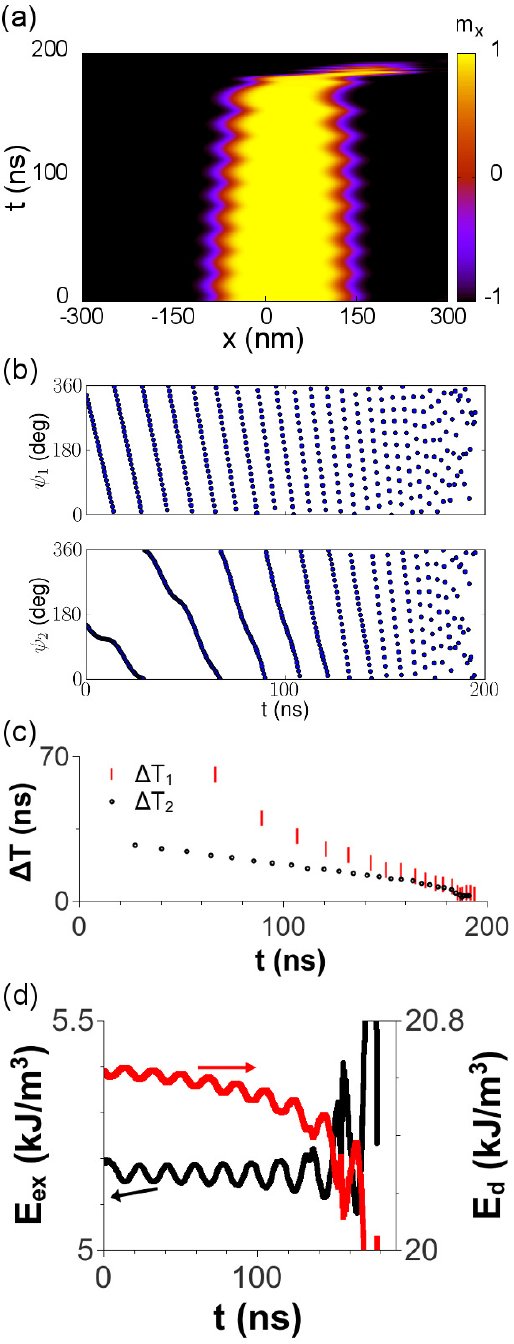}\centering
 \caption{\label{Fig.2} (Color online) (a) Density plot of the average magnetization ($m_x$) along the nanowires axis showing a metastable state of two DWs. The plot of the angular variation of the magnetization in the $yz$ plane ($\psi$ angle in degrees) of the two DWs is shown in (b). (c) Instantaneous period of the two signals from (b). In the region of synchronization the ratio of periods remains constant. (d) Time variation of exchange and demagnetizing energies of (a).}
\end{figure}


The DWs in the nanowire were obtained by different methods: the DWs were nucleated at the ends of the nanowire and subsequently displaced to the center by a spin polarized current, or by using a predefined multi-domain magnetic state. In experiments, DWs are usually injected from nucleation pads at the ends of the nanowire\cite{Glathe} or at the corners of a U-shaped nanowire\cite{Faulkner}.  

Fig.~\ref{Fig.1}(a) shows a snapshot of a pair of TDWs of opposite chirality (HH and TT) in a cylindrical nanowire. The configuration was obtained for a Nickel nanowire with 60nm diameter and 600nm in length (4nm cell size, $\mu_0$M$_s$ = 0.6T, zero crystalline anisotropy).  The transverse DW is the natural ground state. For this diameter, the difference in total micromagnetic energy between the TW and VW is only of 0.6\% of E$_0$, with E$_0 = \mu_0 M_s^2$/2. The form of the walls is triangular and similar with the one of flat nanomagnets\cite{Thiaville}. Fig.~\ref{Fig.1}(b) shows the magnitude (norm) of the divergence of magnetization on a planecut through the center of the nanowire for the snapshot in panel (a). Light areas indicate the maximum values of $\|\nabla\textbf{m}\|$ with the polarity (poles) denoted by $\pm$. We also indicate the polarity of the surface density $\sigma = \pm\textbf{m}$ at the two edges, considering a nanowire uniformly magnetized in the $x$ direction.

The dynamics of a pair of TDW is shown in Fig.~\ref{Fig.2}(a). The figure shows the density plot of the magnetization ($m_x$) along the axis of the nanowire ($x$) during 200ns. The DWs position oscillates while the magnetization inside the DWs precesses out of phase in the $yz$ plane (normal to the cylinder axis) as shown in the plot of the azimuthal angle ($\psi$ in degrees) in Fig.~\ref{Fig.2}(b) for each DW. The precession of the magnetization in the DWs, that is anticlockwise has a variable angular velocity. We observe a metastable bound state with a lifetime of 200ns. Initially, the magnetization in the two DWs is rotating with different angular velocity and after some time the two DWs turn synchronously and annihilate. To determine the instant of synchronization, the instantaneous phases for the two signals in panel (b) were computed using the Hilbert transform. Afterwards, the instantaneous period (frequency) was calculated and is represented in Fig.~\ref{Fig.2}(c). When the two signals synchronize, the ratio of the frequencies remains constant. We observe that the two DWs synchronize after 170ns and shortly after they annihilate creating a small burst of spin waves.

As observed in the Fig.~\ref{Fig.1}(b), the magnetic charge distribution for a TDW is asymmetric, the wide side being a region of high charge concentration\cite{Petit,Zeng}. Precession of the DW magnetization means the precession of the wide side of each DW and therefore of the respective magnetic charge and dipolar moment. The DWs magnetization can point in every direction in the $yz$ plane therefore there is a continuous precession of the DWs charge and of their mutual interaction. The formation of the metastable bound state of two DWs is a result of the charge oscillations of the DWs in the potential landscape of the nanowire (with edges). As shown in Fig.~\ref{Fig.2}(d), its motion is determined by the conversion between the exchange and demagnetizing energies which oscillate in time. The total energy is actually decreasing slowly do to the small damping factor. The dissipation damps the oscillation of the bound state as can be seen in the Fig.~\ref{Fig.2}(a), where after 200ns the pair shrinks and annihilate. At small distances between the TDWs, the exchange energy increases being of the same order as the dipolar energy and tries to align the magnetization of the DWs. 

\begin{figure}[!t]
\center
  \includegraphics[width=6cm]{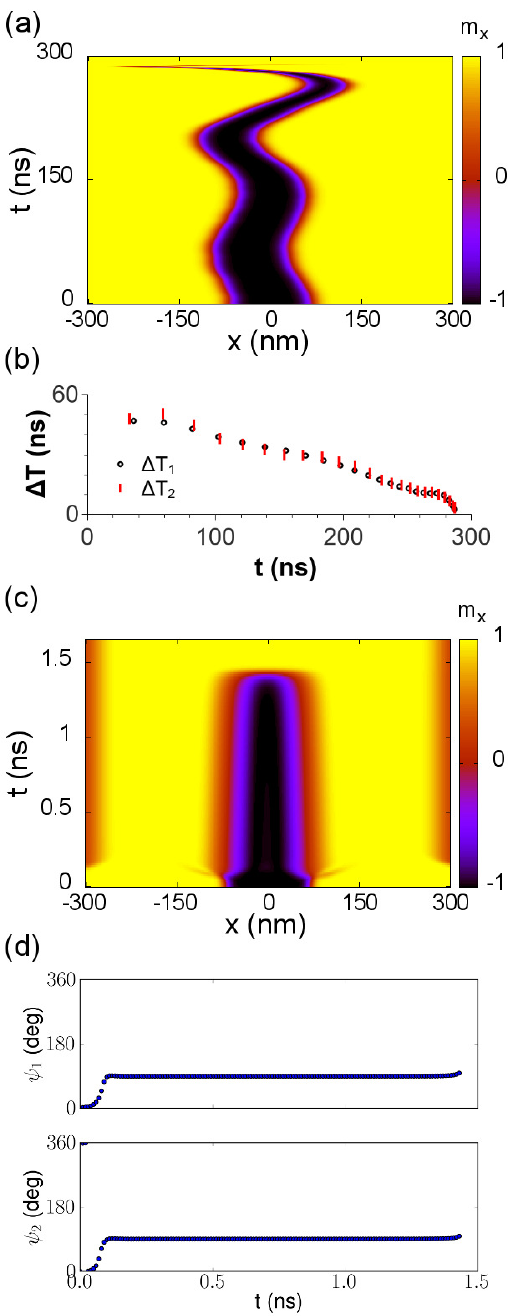}
 \caption{\label{Fig.3} (Color online) Density plot of the average $x$ component of the magnetization, along the nanowire axis, for a cylinder with a diameter of 20nm and length 600nm with no anisotropy (a) and with high transverse ($z$ axis) anisotropy (c). When no anisotropy is present, the DWs form a metastable state and they synchronize after 260ns as shown by the instantaneous period of the two DWs in (b). When high transverse anisotropy is present, the two DWs don't oscillate as theirs azimuthal angles stay constant as observed in (d).}
\end{figure} 


The mutual interaction of the DWs can also be explained by topological considerations. The DWs can be considered as composite objects formed of edge defects with half-integer winding numbers\cite{Tchernyshyov,Kunz} (n = $\pm$1/2). Two edge defects of opposite winding number or opposite skyrmion charge\cite{Tretiakov} annihilate (attract) while two defects of same winding number or same skyrmion charge repel each other. The bound state is therefore a result of the precession of the topological charge around the axis of the nanowire.   

To clearly demonstrate that metastable states exist in cylindrical nanowires, we compared the same initial predefined state of two DWs situated in two identical nanowires (length 600nm, diameter 20nm): one with high uniaxial perpendicular anisotropy and one without anisotropy (as before). When no anisotropy is present (Fig.~\ref{Fig.3}(a)), the magnetization precess freely on the $yz$ plane and various states can appear as seen above. The two DW form a metastable state with a lifetime of 300ns. As the pair width is smaller than in larger diameter nanowires (smaller DW width), the DWs can travel longer distances before reaching the edge. As the pair synchronizes (after 260ns, see panel (b)) and shrinks, the angular velocity and therefore the speed of the pair increases and the amplitude of the spatial oscillation increases before annihilation. 

In the case of the high perpendicular (along $z$ axis) anisotropy (Fig.~\ref{Fig.3}(c)), the magnetization is frustrated and the anisotropy breaks the rotational symmetry. Using a value for the anisotropy of K$_{\perp}$ = 5.2$\times 10^5$J/m$^3$ (value corresponding to Co), we observe that the magnetization is strongly frustrated in the $yz$-plane and it does not precess similar to what happens for DWs in flat nanowires. Pairs of TDWs can form, with the magnetization in the walls pointing in the transverse hard-axis direction (Fig.~\ref{Fig.3}(d)) but no oscillatory bound state appears. The pair of DWs annihilate rapidly after 1.5ns. The anisotropy energy is much higher than the other energies at play and damps strongly the motion as the precession is obstructed.
 
The formation of the metastable bound state of DWs depends on the potential landscape of the sample. If the magnetization of DWs turns asynchronously and is not frustrated, the metastable state can form even in smaller diameter or longer nanowires (not shown). The key conditions for the formation of a bound state in our simulations is the precession speed of each DWs magnetization and the position of the DWs as the tails of the walls are confined by the finiteness of the sample\cite{Dolocan2} and by each other. As long as the above conditions are fulfilled, even in a more complex potential landscape, oscillating metastable pairs appear in nanowires.

\begin{figure}[!t]
\center
  \includegraphics[width=7cm]{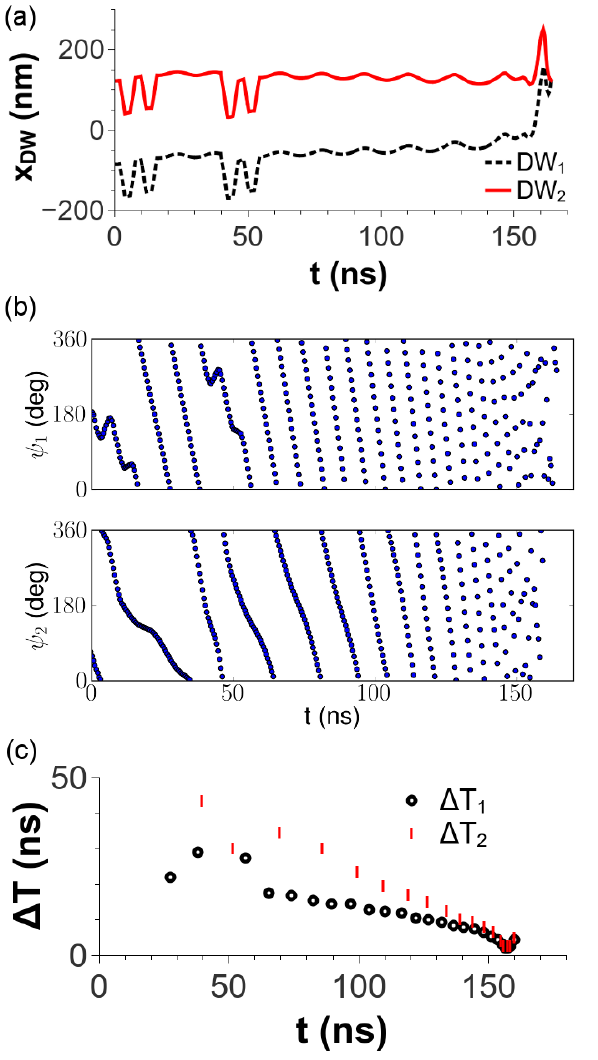}
 \caption{\label{Fig.4} (Color online) (a) Variation of position of two DWs in the forced regime for the same cylinder as in Fig.~\ref{Fig.2}. Four successive pulses of spin polarized current with amplitude of 5$\times10^{11}$A/m$^2$ were applied during 2ns at 2ns intervals in both directions along the $x$-axis. After 25ns, four other pulses where again applied to move the pair of DWs. (b) Angular variation ($\psi$ angle) of the two DWs shown in panel (a). (c) Instantaneous period of the two signals from (b) showing synchronization after 150ns.}
\end{figure}

\textit{Forced regime}. The forced regime covers the interaction of DWs with an external force like a spin polarized current or a magnetic field. We only consider here the former case. Nanosecond dc current pulses were applied in both directions along the symmetry axis of the nanowire ($x$ axis). In all the cases, the DWs move linearly in the direction of the electron flow. This interaction is detailed in Fig.~\ref{Fig.4}(a) where the position of a pair of DWs is shown: eight 2ns dc current pulses are applied with an amplitude of 5$\times10^{11}$A/m$^2$ ($u=42m/s$) to the same initial configuration as in Fig.~\ref{Fig.2}. The damping parameter $\alpha$ is taken as 0.015, while the nonadiabatic parameter $\beta$ = 0.05. The first pulse is applied in the +$x$ direction, while the second pulse is applied in reverse direction, to move the pair back. The delay between the pulses is 2ns. After another 2ns, the sequence is repeted. The DW pair is then relaxed for an initial period of 25ns, after which the same procedure of four pulses as before is applied. Afterward, the pair relaxes freely. The DW pair oscillates during the two relaxation periods and after 170ns annihilates, as the DWs precess synchronously (see the instantaneous periods in Fig.~\ref{Fig.4}(c)). The choice of the amplitude of the current in simulations is made as to have a reasonable speed and exclude nonlinear effects. The pulse duration was deliberately kept short not to nucleate DWs at the ends of the cylinder.

The influence of the current manifests through the modification of the speed and angular velocity of the DW pair during the application of the pulse. The DWs move in the same direction contrary to the case of an external applied field and maintain a constant distance while moving. The current modifies the potential landscape of the nanowire and slows the precession velocity (Doppler shift). As can be observed in Fig.~\ref{Fig.4}(b), the rotation direction and angular velocity can change due to changes in the torque applied by the current. The uniform precession is disturbed and the main interaction of the DWs is with the current. The pair width remains constant during the application of the current and after the extinction of the pulse the DWs restart to synchronize. Therefore, the DWs pair can be moved long distances with current pulses without disrupting the bound state and restarts to oscillate when the current is turned off.

The time evolution of the DWs can be explained with the 1D analytical model. From Eq.~\eqref{eq1}, the velocity and the precession speed of the TW in cylindrical nanowires can be calculated for the case $\alpha\neq\beta$ (with H$_{eff}$=0) as: $v  = \frac{1+\beta\alpha}{1+\alpha^2}u $ and $\dot{\phi}  = \frac{\beta - \alpha}{1+\alpha^2}\frac{u}{\Delta}$. The speed of the DW is almost linear ($\alpha, \beta \ll 1$) in $u$ while the precession depends on the sign of $\alpha - \beta$ (here $<<$ 0) and $u$. In the case of small cylindrical nanowires, no threshold current is needed to move the DW and the influence of the current is determined by the above expressions. The numerically calculated velocity (from Fig.~\ref{Fig.4}) corresponds to the analytical value. Changing the nonadiabatic parameter $\beta$ to values higher or smaller than $\alpha$, modifies the precession speed of both DW and thus does not affect the metastable boundstate.

In real nanowires, the DWs are usually pinned by defects with an external force (or just thermal force) needed to depin them. The DWs can still precess and the bound state can form if two DWs are close enough and precess out of phase. The movement of one DW can be estimated from Eq.~\eqref{eq1} with a pinning energy term as for artificial defects\cite{Martinez,Gonzalez}. The influence of temperature should not affect the apparition of bound states as the thermal energy at room temperature is only 25meV. Although the precession in the two DW should be affected in the same manner, the thermal energy can influence the DW that is closer to the edge if this one is at the top of the potential curve (Fig.~\ref{Fig.1}(c)) and break the pair. The lifetime of the bound states could be diminished. 



In summary, in this paper we reported on the metastable DW states that appear in ideal small diameter cylindrical nanowires. We showed that in special conditions the DW pairs form metastable bound states that can be local equilibrium states with a lifetime longer than 200ns. These states oscillate with small amplitude around theirs position. Theirs presence can perturb the performance of devices based on DW motion. One way to stabilize the DWs position is through strong anisotropy or pinning by patterning of notches. We also observed that the DW pairs are stabilized by an external force, as a spin polarized current, and can be moved over long distances. If hard-axis anisotropy is present, it frustrates the precession of the magnetization and the motion of DWs under current. 


The author wish to thank L. Raymond for the access to the Lafite servers and is grateful for the support of the NANOMAG platform by FEDER and Ville de Marseille. Part of the computations were performed at the Mesocentre d'Aix-Marseille University.



\end{document}